\documentstyle{mn}

\input epsf

\begin{document}

\def\Ha{H$\alpha\ $}
\def\Hb{H$\beta\ $}
\def\Lya{Ly$\alpha\ $}
\def\Lyb{Ly$\beta\ $}
\def\Lyg{Ly$\gamma\ $}
\def\Lyd{Ly$\delta\ $}
\def\Lye{Ly$\epsilon\ $}
\def\Lyz{Ly$\zeta\ $}
\def\Lyet{Ly$\eta\ $}
\def\Lyth{Ly$\theta\ $}
\def\LCDM{$\Lambda$CDM\ }
\def\HI{\hbox{H~$\rm \scriptstyle I\ $}}
\def\HII{\hbox{H~$\rm \scriptstyle II\ $}}
\def\DI{\hbox{D~$\rm \scriptstyle I\ $}}
\def\HeI{\hbox{He~$\rm \scriptstyle I\ $}}
\def\HeII{\hbox{He~$\rm \scriptstyle II\ $}}
\def\HeIII{\hbox{He~$\rm \scriptstyle III\ $}}
\def\CII{\hbox{C~$\rm \scriptstyle II\ $}}
\def\CIII{\hbox{C~$\rm \scriptstyle III\ $}}
\def\CIV{\hbox{C~$\rm \scriptstyle IV\ $}}
\def\NV{\hbox{N~$\rm \scriptstyle V\ $}}
\def\OIII{\hbox{O~$\rm \scriptstyle III\ $}}
\def\OIV{\hbox{O~$\rm \scriptstyle IV\ $}}
\def\OVI{\hbox{O~$\rm \scriptstyle VI\ $}}
\def\SiIV{\hbox{Si~$\rm \scriptstyle IV\ $}}
\def\NHI{N_{\rm HI}}
\def\NHeII{N_{\rm HeII}}
\def\cm2{\,{\rm cm$^{-2}$}\,}
\def\kms{\,{\rm km\,s$^{-1}$}\,}
\def\skm{\,({\rm km\,s$^{-1}$})$^{-1}$\,}
\def\kmsmpc{\,{\rm km\,s$^{-1}$\,Mpc$^{-1}$}\,}
\def\hmpc{\,h^{-1}{\rm \,Mpc}\,}
\def\mpch{\,h{\rm \,Mpc}^{-1}\,}
\def\hkpc{\,h^{-1}{\rm \,kpc}\,}
\def\ev{\,{\rm eV\ }}
\def\kel{\,{\rm K\ }}
\def\intunits{\,{\rm ergs\,s^{-1}\,cm^{-2}\,Hz^{-1}\,sr^{-1}}}
\def\ltsima{$\; \buildrel < \over \sim \;$}
\def\lsim{\lower.5ex\hbox{\ltsima}}
\def\gtsima{$\; \buildrel > \over \sim \;$}
\def\gsim{\lower.5ex\hbox{\gtsima}}
\def\etal{{ et~al.~}}
\def\aj{AJ}
\def\ana{A\&A}
\def\apj{ApJ}
\def\apjs{ApJS}
\def\mn{MNRAS}

\journal{Preprint-00}

\title{Colour corrections for high redshift objects due to intergalactic attenuation}

\author[A. Meiksin]{Avery Meiksin \\
SUPA\thanks{Scottish Universities Physics Alliance},
Institute for Astronomy, University of Edinburgh,
Blackford Hill, Edinburgh\ EH9\ 3HJ, UK}

\pubyear{2005}

\maketitle

\begin{abstract}
Corrections to the magnitudes of high redshift objects due to
intergalactic attenuation are computed using current estimates of the
properties of the intergalactic medium. The results of numerical
simulations are used to estimate the contributions to resonant
scattering from the higher order Lyman transitions. Differences of
$0.5-1$ magnitude from the previous estimate of Madau (1995) are
found. Intergalactic $k_{\rm IGM}$-corrections and colours are
provided for high redshift starburst galaxies and Type I and Type II
QSOs for several filter systems used in current and planned deep
optical and infra-red surveys.
\end{abstract}

\begin{keywords}
galaxies:\ high redshift - galaxies:\ photometry - intergalactic medium - quasars:\ absorption lines - quasars:\ general - surveys
\end{keywords}

\section{Introduction} \label{sec:introduction}

Over the past decade, deep optical and infra-red surveys have enabled
giant strides to be taken in elucidating the nature and properties of
objects that populate the high redshift universe. The band-dropout
method has unveiled a population of Lyman break galaxies at
$z\approx3$ (Guhathakurta \etal 1990; Bithell 1991; Steidel \&
Hamilton 1992, 1993; Steidel, Pettini \& Hamilton 1995; Steidel \etal
2003) and higher (Sawicki \& Thompson 2005). The selection method was
successfully applied to the Hubble Deep Field (HDF) (Giavalisco,
Steidel \& Macchetto 1996; Steidel \etal 1996), broadening the
redshift range and volume coverage over previous surveys. Most
recently the Ultra Deep Field (UDF) was exploited to discover objects
as distant as $z\gsim6$ (Stanway, Bunker \& McMahon 2003; Bouwens
\etal 2004; Yan \& Windhorst 2004; Giavalisco \etal 2004).

An alternative selection method for identifying high redshift objects
relies on combinations of broadband colours to estimate photometric
redshifts, for which the most likely redshift is assigned based on
predicted spectral energy distributions (Sawicki, Lin \& Yee 1997;
Csabai \etal 2000).

The modelling of the high redshift objects through population
synthesis, applied to a combination of spectroscopic data and
broad-band colours, suggests that most of the high redshift objects
are star-forming galaxies (Madau \etal 1996; Metcalfe \etal 2001;
Papovich, Dickinson \& Ferguson 2001; Pettini\etal 2001; Shapley \etal
2003). A few of the Lyman break galaxies contain Active Galactic
Nuclei (Steidel \etal 2002), used to determine the faint end of the
QSO luminosity function at $z\approx3$ (Hunt \etal 2004).

Parallel to these surveys have been several searches for high redshift
Quasi-Stellar Objects (QSOs). The discovery of a few dozen $z>3.6$
QSOs by the Sloan Digital Sky Survey (SDSS) has made possible a new
evaluation of the bright end of the QSO luminosity function and its
evolution at these high redshifts (Fan \etal 2001). The results are
currently being revised (Richards \etal 2005) based on the much larger
numbers now detected, including over 500 at $z>4$ (Schneider \etal
2005).

Similar surveys are expected to continue well into the future,
including some now in progress, such as the Canada-France-Hawaii
Telescope Legacy Survey (CFHTLS\footnote {{\rm
www.cfht.hawaii.edu/Science/CFHLS/ }}) and the UKIRT Infrared Deep Sky
Survey (UKIDSS\footnote{{\rm www.ukidss.org }}; Hewett \etal, in
preparation), or planned for telescopes currently under development,
such as the Visible \& Infrared Survey Telescope for Astronomy
(VISTA\footnote{ {\rm www.vista.ac.uk }}) and the {\it James Webb
Space Telescope} (JWST\footnote{{\rm www.ngst.nasa.gov }}).

Crucial to all these analyses is an accurate estimate of the amount of
intergalactic attenuation due to intervening absorption systems. Most
have relied on the standard work of Madau (1995), whose assessment was
based on the then most current understanding of the properties and
distributions of intervening systems. Madau (1995) estimated the
blanketing due to the resonant (Lyman series) scattering of photons
assuming idealised forms for the \HI column density distribution of
the absorbers. The contributions of the different orders in the Lyman
series depend on the full line-shape of the absorber, precluding a
direct scaling of line-centre optical depths based on pure atomic
physics considerations. Instead the broadening of the absorbers must
be included, the distribution of which has since been shown to be
sensitive to column density (Kirkman \& Tytler 1997; Kim, Cristiani \&
D'Odorico 2002b). Madau (1995) adopted a constant Doppler parameter
for all absorption systems, varying the value to probe the sensitivity
of the total amount of attenuation to this variable. A blanketing
formalism based on Poisson placement of the absorbers was used to
predict the effective optical depths, although in principle small
scale clustering of the absorption systems will affect the total
amount of blanketing, and such correlations have been detected
(Kirkman \& Tytler 1997; Kim \etal 2002a).

Since Madau's seminal work, numerical simulations of the intergalactic
medium have yielded results matching the measured distributions of the
\Lya flux distributions to an accuracy of a few percent, as well as
the line parameters (allowing for extra heating) (Meiksin, Bryan \&
Machacek 2001). Numerical simulations have also reproduced the \HI
column density dependence of the Doppler parameter envelope (Misawa
\etal 2004), as well as correlations in the \HI flux distribution
(Croft \etal 2002; Meiksin \& White 2004). The understanding of the
mean \Lya intergalactic optical depths has improved substantially over
the past decade (see Meiksin \& White 2004 for a summary and Kirkman
\etal 2005 for subsequent results). The simulations contain the
information necessary to extract the contributions from all higher
order transitions to the blanketing. Although simulations do not
recover the full numbers of Lyman Limit Systems observed (Gardner
\etal 1997; Meiksin \& White 2004), an assessment of their numbers
over the redshift range $0.3\lsim z\lsim4$ has been made by
Stengler-Larrea \etal (1995). These improvements now permit a much
more secure determination of the amount of intergalactic attenuation
to be made.

\section{Numerical simulation predictions for intergalactic attenuation}

Attenuation due to intervening intergalactic hydrogen arises through
two principle mechanisms, resonant scattering by Lyman transitions and
photoelectric absorption. Additional contributions are made by
intervening metal systems and intervening helium. The contributions of
metals and \HeI are small (Madau 1995), while \HeII will contribute
only at wavelengths $\lambda<228(1+z)$\AA\ for a source at redshift
$z$.

The hydrogen attenuation values are based on the simulation of a \LCDM
model by Meiksin \& White (2004) with $\Omega_M=0.3$, $\Omega_v=0.7$,
$h=0.7$, $n=0.95$, and normalised to $\sigma_8=0.92$, consistent with
{\it WMAP} constraints. The simulation used a pure particle mesh (PM)
scheme (Meiksin \& White 2001), mimicking the temperature of the gas
using a polytropic equation of state and assuming the gas and dark
matter have the same spatial distribution. The mean transmitted \Lya
fluxes are normalised using the values listed in Table~3 of Meiksin \&
White (2004). The attenuation values resulting from higher order Lyman
transitions are computed directly from the simulation results.

The mean optical depth value for the Lyman transition $n\rightarrow1$
is defined by
\begin{equation}
\bar\tau_n\equiv-\ln\langle\exp(-\tau_n)\rangle,
\label{eq:taun}
\end{equation}
where $\langle\exp(-\tau_n)\rangle$ is the corresponding mean
transmitted flux. (Here, the convention $\tau_\alpha=\tau_2$,
$\tau_\beta=\tau_3$, etc. will be adopted.)  For $2 < z < 4$, the
measured values of the mean \Lya transmitted flux are fit to within
2\% using
\begin{equation}
\bar\tau_\alpha = 0.00211(1+z)^{3.7}\qquad (z<4).
\label{eq:tauLya1}
\end{equation}
The sharp reduction in flux at higher redshifts is adequately fit by
\begin{equation}
\bar\tau_\alpha = 0.00058(1+z)^{4.5}\qquad (z>4).
\label{eq:tauLya2}
\end{equation}
For $z>6$, the measured values become quite uncertain, but the
objects become dimmed by 4 magnitudes or more so that accurate
values will normally not be required.

\begin{table}
\caption{Ratio of $\bar\tau_n/\bar\tau_\alpha$ for Lyman transition
$n\rightarrow1$ for $n=3$ to 9. Each transition contributes at
wavelength $\lambda$ for an object at redshift $z$ only under the
condition $z_n<z$, where $z_n\equiv(\lambda/\lambda_n)-1$ and
$\lambda_n$ is the (rest) wavelength of the transition.
}
\begin{center}
\begin{tabular}{|c|l|} \hline
\hline 
$n$ & $\bar\tau_n/\bar\tau_\alpha$ \\
\hline
3 & $0.348[0.25(1+z_n)]^{1/3},\, (z_n < 3); 0.348[0.25(1+z_n)]^{1/6},\, (z_n>3)$ \\
4 & $0.179[0.25(1+z_n)]^{1/3},\, (z_n < 3); 0.179[0.25(1+z_n)]^{1/6},\, (z_n>3)$ \\
5 & $0.109[0.25(1+z_n)]^{1/3},\, (z_n < 3); 0.109[0.25(1+z_n)]^{1/6},\, (z_n>3)$ \\
6 & $0.0722[0.25(1+z_n)]^{1/3}$ \\
7 & $0.0508[0.25(1+z_n)]^{1/3}$ \\
8 & $0.0373[0.25(1+z_n)]^{1/3}$ \\
9 & $0.0283[0.25(1+z_n)]^{1/3}$ \\
\hline
\end{tabular}
\end{center}
\label{tab:taun-to-taua}
\end{table}

The higher order terms in the Lyman series are characterised by the
ratio $\bar\tau_n/\bar\tau_\alpha$ to factor out most of the redshift
dependence. A residual dependence, however, remains. These are
accurately fit (to within a few percent) by weak powers of
$(1+z)$. The ratios for the first 7 orders (\Lyb to \Lyth) after \Lya
are provided in Table~\ref{tab:taun-to-taua}. Higher orders ($n>9$)
are found to scale very nearly according to the atomic physics
prediction for line-centre optical depth values:
\begin{equation}
\frac{\bar\tau_n}{\bar\tau_\theta}\approx\frac{720}{n(n^2-1)}.
\label{eq:tauiscale}
\end{equation}
Terms up to $n=31$ are included:\ higher order Lyman series
transitions contribute negligibly.

The contribution due to photoelectric absorption is split into two
parts, the contribution from systems optically thin at the Lyman edge
and the contribution from Lyman Limit Systems. The contribution from
optically thin systems is given in the linear approximation
$\tau_L=r/r_0$ (Zuo 1992), where $r$ is the proper distance between
the emitting object at redshift $z$ and the redshift
$z_L=\lambda/\lambda_L - 1$, where $\lambda_L$ is the wavelength at
the Lyman edge ($\lambda_L=912$\AA), $\lambda$ is the observed
wavelength, and $r_0$ is the attenuation length of photons at the
photoelectric edge. Using the result for the attenuation length from
Meiksin \& White (2004), this is well-approximated by
\begin{equation}
\tau_L^{\rm IGM}=0.805(1+z_L)^3\left(\frac{1}{1+z_L}-\frac{1}{1+z}\right).
\label{eq:tauL_IGM}
\end{equation}

The contribution due to Poisson-distributed Lyman Limit Systems is
given by (Zuo 1992)
\begin{equation}
\tau_L^{LLS}=\int_{z_L}^z dz'\int_1^\infty d\tau_L\, \frac{\partial^2 N}
{\partial\tau_L \partial z'}\left\{1-\exp\left[-\tau_L\left(\frac{1+z_L}{1+z'}
\right)^3\right]\right\},
\label{eq:tauL_LLS}
\end{equation}
where $\partial^2 N/\partial\tau_L \partial
z=A\tau_L^{-\beta}(1+z)^\gamma$ is assumed, corresponding to a number
density $dN/dz=N_0(1+z)^\gamma$ for systems with $\tau_L>1$, where
$N_0=A/(\beta-1)$. The integral has the power series solution
\begin{eqnarray}
\tau_L^{LLS} = & \frac{N_0}{4+\gamma-3\beta}\left[\Gamma(2-\beta,1)-e^{-1}
-\Sigma_{n=0}^\infty\frac{\beta-1}{n+1-\beta}\frac{(-1)^n}{n!}\right] \nonumber \\
&\times\left[(1+z)^{-3(\beta-1)+\gamma+1}\left(\frac{\lambda}{\lambda_L}
\right)^{3(\beta-1)}-\left(\frac{\lambda}{\lambda_L}\right)^{\gamma+1}
\right] \nonumber \\
&-N_0\Sigma_{n=1}^\infty\frac{\beta-1}{(3n-\gamma-1)(n+1-\beta)}
\frac{(-1)^n}{n!} \nonumber \\
&\times \left[(1+z)^{\gamma+1-3n}\left(\frac{\lambda}{\lambda_L}\right)^{3n}-
\left(\frac{\lambda}{\lambda_L}\right)^{\gamma+1}\right].
\label{eq:tauLLLS}
\end{eqnarray}
The first 10 terms of each series provide a high level of
convergence. The results are normalised to $N_0=0.25$, $\beta=1.5$ and
$\gamma=1.5$ (Stengler-Larrea \etal 1995). Although the result of
Stengler-Larrea \etal applies only for $z\lsim4$, the redshift dependence
is extrapolated to higher redshifts here, noting that the contribution
of these systems is uncertain at these redshifts.

\begin{figure}
\begin{center}
\leavevmode \epsfxsize=3.3in \epsfbox{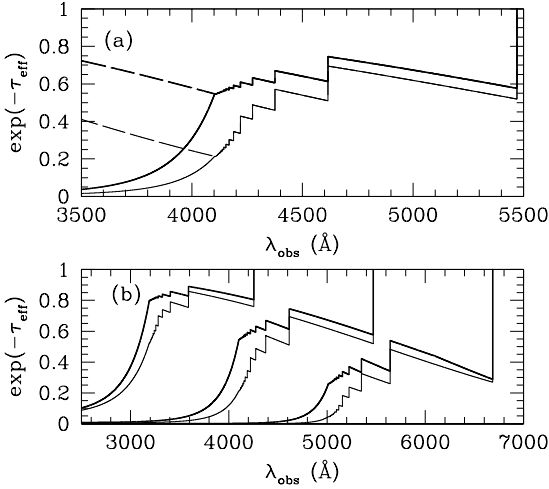}
\end{center}
\caption{Intergalactic transmission as a function of observed
wavelength. (a) Mean intergalactic transmission for a source at
redshift $z=3.5$ (solid lines). The mean intergalactic transmission
assuming no Lyman Limit Systems lie along the line-of-sight is shown
by the dashed lines. (b) Mean intergalactic transmission for, from
left to right, sources at redshifts $z=2.5$, 3.5 and 4.5.  The heavy
lines are the estimates from this paper. In both panels, the light
lines are the estimates from Madau (1995).
}
\label{fig:tauev}
\end{figure}

The total optical depth is the sum of the resonant and photoelectric
contributions. The photoelectric contribution is dominated by sytems
with $\tau_L\approx1$; optically thin sytems contribute only a small
amount while very optically thick systems are too few in number to
contribute much on average. While the photoelectric contribution due
to the optically thin IGM is well-determined due to the large number
of optically thin systems, the contribution from Lyman Limit Systems
is highly variable, depending on the chance that a Lyman Limit System
lies along the relevant path length. For instance, for a source at
$z=3.5$, the Lyman limit lies at the observed wavelength
$\lambda=4104$\AA. The number of Lyman Limit Systems that will fall at
a redshift corresponding to the $U_n$-band wavelength range of
approximately 3000\AA$-$4000\AA\ is 2.1. The Poisson probability that
no Lyman Limit System lies in this range is then $e^{-2.1}=0.12$. This
will introduce large fluctuations in the amount of attenuation at
these short wavelengths (Zuo \& Phinney 1993; Madau 1995), a fact
which should be borne in mind when interpreting the number of source
detections in Lyman dropout surveys.

\begin{table*}
\centering
\begin{minipage}{180mm}
\caption{Intergalactic \HI transmission as a function of observed wavelength.}
\begin{tabular}{@{}lrrrrrrrrrrrr@{}}
\hline\hline
$\lambda$ (\AA) & $z=1.5$& $z=2.0$& $z=2.5$& $z=3.0$& $z=3.5$& $z=4.0$& $z=4.5$& $z=5.0$& $z=5.5$& $z=6.0$& $z=6.5$& $z=7.0$\\
\hline
1730.0 & 0.323241 & 0.150055 & 0.074558 & 0.038532 & 0.020407 & 0.010983 & 0.005976 & 0.003277 & 0.001807 & 0.001001 & 0.000556 & 0.000310\\
1731.0 & 0.323538 & 0.150066 & 0.074510 & 0.038482 & 0.020369 & 0.010956 & 0.005958 & 0.003266 & 0.001800 & 0.000996 & 0.000553 & 0.000308\\
1732.0 & 0.323836 & 0.150077 & 0.074462 & 0.038433 & 0.020330 & 0.010929 & 0.005940 & 0.003254 & 0.001793 & 0.000992 & 0.000550 & 0.000306\\
1733.0 & 0.324135 & 0.150089 & 0.074414 & 0.038383 & 0.020292 & 0.010902 & 0.005922 & 0.003243 & 0.001785 & 0.000987 & 0.000548 & 0.000304\\
1734.0 & 0.324436 & 0.150101 & 0.074367 & 0.038334 & 0.020253 & 0.010875 & 0.005905 & 0.003231 & 0.001778 & 0.000983 & 0.000545 & 0.000303\\
\hline
\end{tabular}
\begin{tabular}{@{}l@{}}
Note:\ The full table is published in the electronic version of the paper. A portion is shown here only for guidance regarding its form and content.
\end{tabular}
\end{minipage}
\label{tab:trans}
\end{table*}

An indication of the level of contribution of Lyman Limit Sytems to
the total optical depth is provided in Figure~\ref{fig:tauev}a, which
shows the level of intergalactic transmission as a function of
observed wavelength and the transmission with the contribution of
Lyman Limit Systems removed. The total intergalactic transmissions
allowing for \HI attenuation are shown as a function of the observed
wavelength for sources at $z=2.5$, 3.5 and 4.5 in
Figure~\ref{fig:tauev}b, and in Table~\ref{tab:trans}.  The figures
include comparisons with the corresponding estimates of Madau (1995),
which tend to lower transmission levels, primarily as a result of
differences in the estimates of the contributions of resonant
absorption.

\section{Effect of intergalactic attenuation on broadband magnitudes}

\begin{figure}
\begin{center}
\leavevmode \epsfxsize=3.3in \epsfbox{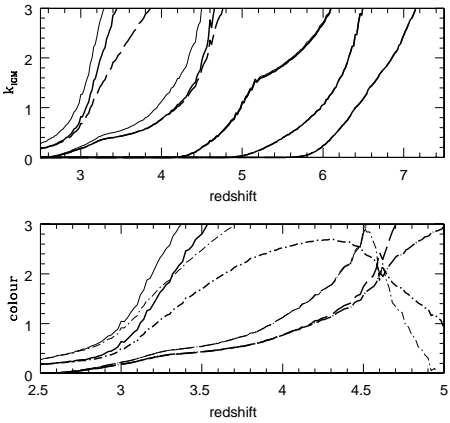}
\end{center}
\caption{Intergalactic attenuation $k_{\rm IGM}$-correction and
colours for a 600~Myr starburst with continuous star formation. (a)\
From left to right, $k_{\rm IGM}$ for $U_n$, $G$, ${\cal R}$, $I$ and
$z^\prime$ bands (solid lines). For the $U_n$ and $G$ bands, $k_{\rm
IGM}$ assuming the absence of Lyman Limit Systems is shown by the
dashed lines. The absence of Lyman Limit Systems leaves the results
for the remaining bands unaltered. The light solid lines correspond to
the predictions for $k_{\rm IGM}$ using the attenuation model of Madau
(1995). (b)\ The $U_n-G$ (solid) and $G-{\cal R}$ (long dashed)
colours of the starburst. The dotted short-dashed lines and dotted
long-dashed line show $U_n-G$ and $G-{\cal R}$, respectively,
assuming the absence of Lyman Limit Systems. The light lines
correspond to the predictions based on the attenuation model of Madau
(1995).
}
\label{fig:UnGR-Iz_sb600}
\end{figure}

In the AB-magnitude system, the apparent magnitude of a source of
intrinsic flux $f_\nu$ (in cgs units) measured through a filter with
(normalised) transmissivity $T(\nu)$ is given by
\begin{equation}
m_{AB}=-2.5\log_{10}\int d\nu\, f_\nu \exp(-\tau_{\rm eff}) T(\nu) - 48.59.
\label{eq:mAB}
\end{equation}
The difference between the magnitudes with and without intergalactic
attenuation will be designated as the intergalactic k-correction
$k_{\rm IGM}=m_{AB}(\tau_{\rm eff}) - m_{AB}(\tau_{\rm eff}=0)$.
Because source spectra are typically slowly varying over a bandwidth,
the $k_{\rm IGM}$-correction is fairly independent of the nature of
the source. The exception is when either the source has narrow
features that dominate the light, such as strong emission or
absorption lines, or when the amount of intergalactic attenuation
varies rapidly within a band, reducing the wavelength range of the
source spectrum that contributes to the total magnitude, and so
enhancing the effect of any differences between source spectra.

To indicate the typical role intergalactic attenuation plays on the
colours of high redshift objects, the effect of intergalactic
attenuation is evaluated for four model sources:\ starbursts of ages
3~Myr and 600~Myr, roughly bracketing the range inferred for Lyman
Break Galaxies (Papovich, Dickinson \& Ferguson 2001), and Type I and
Type II QSOs. The starburst spectra assume continuous star-formation
with solar metallicity and a Salpeter Initial Mass Function and were
generated by the STARBURST99 model of Leitherer \etal (1999). The Type
I QSO spectrum is the composite spectrum constructed from over 2200
spectra homogeneously selected from the SDSS QSO survey covering the
redshift interval $0.044<z<4.789$ and restframe wavelength range
800\AA$-$8555\AA\ (Vanden Berk \etal 2001). The Type II QSO spectrum
is based on observations of CXO~52 (Stern \etal 2002), as described in
Meiksin (2005). This Type II QSO was chosen because it has high
equivalent width emission lines that will dominate the magnitude of
the band in which they lie, and so represents a class of objects that
will have unusual broadband colours that vary substantially with
redshift.

\begin{figure*}
\begin{minipage}{150mm}
\leavevmode \epsfxsize=6in \epsfbox{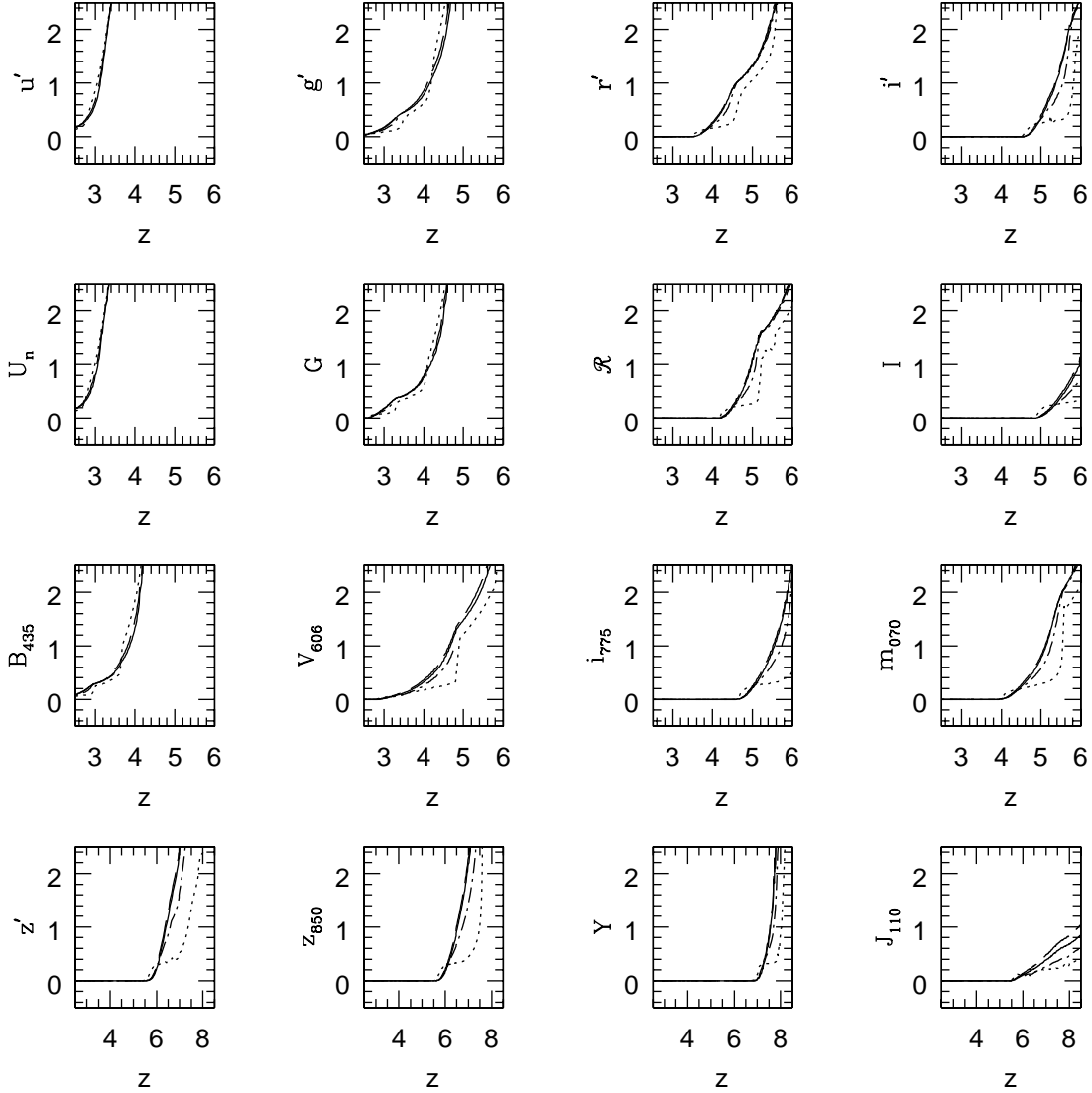}
\caption{Intergalactic attenuation $k_{\rm IGM}$-corrections as a
function of source redshift for a 600~Myr starburst (solid lines), a
3~Myr starburst (long-dashed lines), a Type I QSO (dot-dashed lines)
and a Type II QSO with high equivalent width emission lines (dotted
lines). There is little variation between the sources except for the
Type II QSO because of the presence of strong emission lines. The rows
correspond to the following filter systems. First row:\ Sloan; second
row:\ Steidel $U_nG{\cal R}$ filters and $I$-band; third row:\ {\it
HST} bands $B_{435}$, $V_{606}$, $i_{775}$ and {\it JWST} $m_{070}$;
fourth row:\ Sloan $z'$, {\it HST} $z_{850}$ and $J_{110}$ and UKIDSS
$Y$. The $k_{\rm IGM}$-corrections are negligible for the {\it HST}
$H_{160}$, {\it JWST} $m_{150}$ and UKIDSS $J$, $H$ and $K$ bands for
$z<8.5$.
}
\label{fig:kIGM}
\end{minipage}
\end{figure*}

\begin{figure*}
\begin{minipage}{180mm}
\leavevmode \epsfxsize=8in \epsfbox{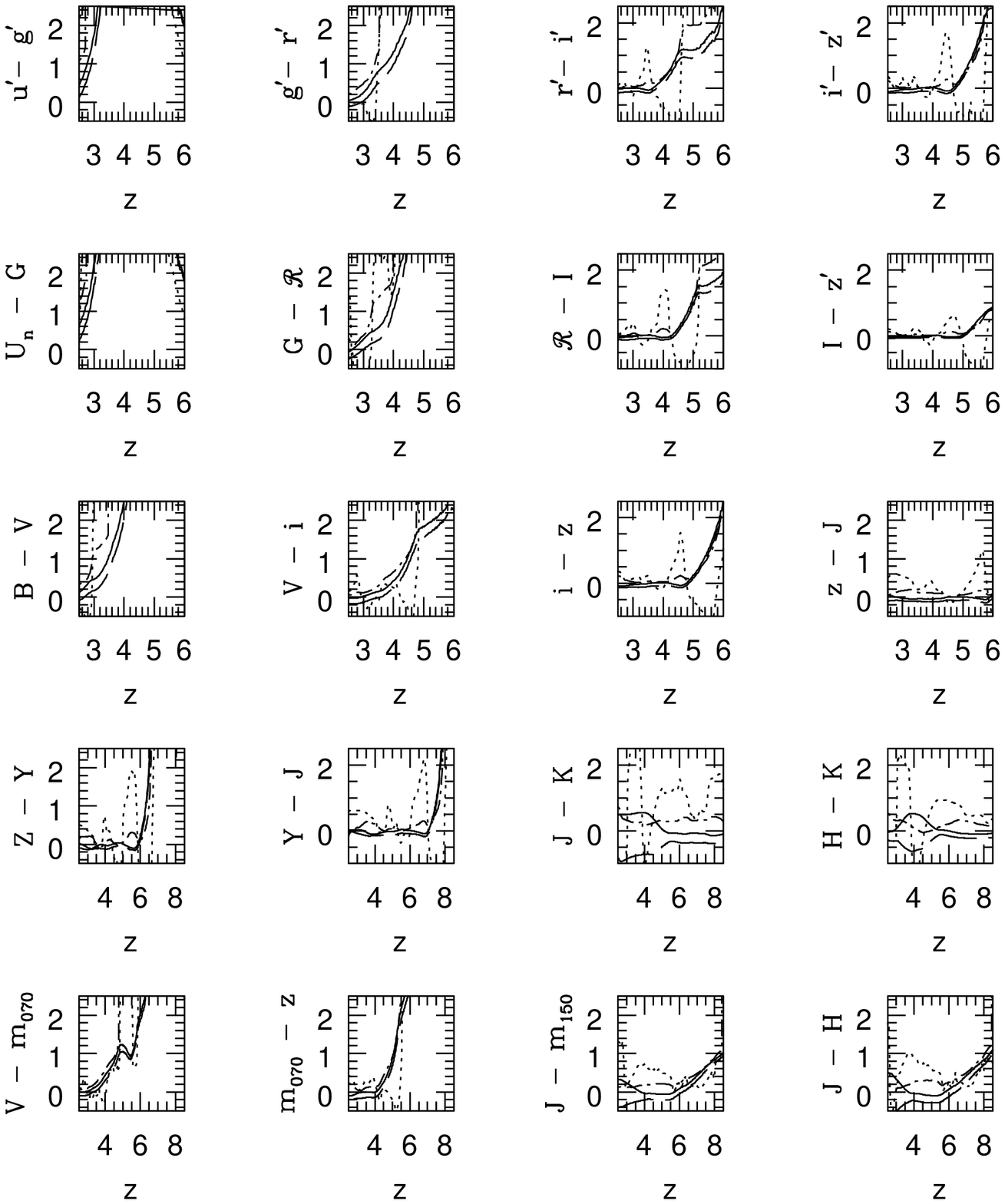}
\caption{Colour predictions including the effects of intergalactic
attenuation for a 600~Myr starburst (solid lines), a 3~Myr starburst
(long-dashed lines), a Type I QSO (dot-dashed lines) and a Type II
QSO with high equivalent width emission lines (dotted lines). In the
third and fifth rows, the $B$, $V$, $i$, $z$, $J$ and $H$ magnitudes
refer to the {\it HST} $B_{435}$, $V_{606}$, $i_{775}$, $z_{850}$,
$J_{110}$ and $H_{160}$ bands, respectively. The fourth row magnitudes
refer to the UKIDSS $ZYJHK$ bands.
}
\label{fig:colours}
\end{minipage}
\end{figure*}

The effect of intergalactic attenuation on broadband magnitudes is
computed for several filter systems relevant to recent and planned
major surveys:\ the Sloan $u^\prime g^\prime r^\prime i^\prime
z^\prime$ system (assuming airmass 1.3 response for a point source);
the $U_nG{\cal R}$ system of Steidel's group, supplemented by the
$I$-band adopting the Harris $I$ filter; the {\it Hubble Space
Telescope} (HST) ACS/NICMOS $B_{435}$ ($F435W$), $V_{606}$ ($F606W$),
$i_{775}$ ($F775W$), $z_{850}$ ($F850LP$), $J_{110}$ ($F110W$) and
$H_{160}$ ($F160W$) bands (including instrumental responses); the
UKIDSS $ZYJHK$ filters; and the {\it JWST} $F070W$ (designated
$m_{070}$ and approximated as a square transmission response) and
$F150W$ (designated $m_{150}$) bands. All magnitudes are computed on
the AB system.

The $k_{\rm IGM}$-corrections and colours for the $U_n$, $G$ and
${\cal R}$ filters of Steidel \& Hamilton (1992) as well as the $I$
and $z^\prime$ bands, are shown in Figure~\ref{fig:UnGR-Iz_sb600} for
the 600~Myr old starburst. The effect of internal redenning is not
included. This will negligibly affect $k_{\rm IGM}$, so the magnitude
corrections to the colours due to redenning will simply be
additive. Also shown are the results assuming no Lyman Limit Systems
intercept the line-of-sight. The mean contribution of Lyman Limit
Systems has a substantial effect on $U_n$, a small effect on $G$, and
a negligible effect on the longer wavelength bands for $k_{\rm
IGM}<3$. A comparison with the attenuation model of Madau (1995) shows
a reduction in $k_{\rm IGM}$ of about 0.5 magnitude, although the
effect on the colours approach a shift of a full magnitude. In terms
of photometric redshifts, the difference in attenuation models
produces a shift of $\Delta z=0.1-0.2$ for $U_n-G$ and $G-{\cal R}$.
The $k_{\rm IGM}$-corrections and various colour combinations for all
the sources considered are shown in Figures~\ref{fig:kIGM} and
\ref{fig:colours}. The $k_{\rm IGM}$-corrections are nearly
independent of the source spectrum, except for the Type II QSO for
which the magnitudes may be dominated by emission lines (Meiksin
2005). The colours also are similar except for the Type II QSO,
although certain colours, such as UKIDSS $J-K$ and $H-K$, are
particularly effective at separating the other sources.

\section{Summary}

The amount of the attenuation of light from high redshift objects on
passing through the IGM has been computed based on recent measurements
of the mean transmitted \Lya flux through the IGM and recent
assessments of the numbers of Lyman Limit Systems. The properties of
the IGM required to compute the attenuation due to resonant Lyman
photon scattering were based on numerical simulations matching the
measured properties of the IGM. Differences from the predictions of
the model of Madau (1995) are found for $k_{\rm IGM}$ and colours of
$0.5-1$ magnitude for bands containing restframe \Lya and shorter
wavelengths.

Intergalactic $k_{\rm IGM}$-corrections are provided for filter
systems used in current or planned deep optical and infra-red surveys,
{\it viz.} the Sloan $u^\prime g^\prime r^\prime i^\prime z^\prime$
system, Steidel $U_n G {\cal R}$ and the $I$-band, UKIDSS $ZYJHK$, the
{\it HST} $B_{435}$, $V_{606}$, $i_{775}$, $z_{850}$, $J_{110}$ and
$H_{160}$ bands, and the {\it JWST} $F070W$ and $F150W$ bands.

Colours based on the above bands are provided for starbursts of ages
3~Myr and 600~Myr, typical of $z\approx3$ Lyman Break Galaxies, and
Type I and Type II QSOs, over the redshift range $2.5<z<8.5$. The
results show which colours are most effective for distinguishing
between different objects and the typical values to expect, which may
be used for the planning and analysis of current and upcoming deep
surveys.

\section*{Acknowledgments}

The author thanks Dr Steve Warren for kindly providing total response curves
for the UKIDSS filter system.


\end{document}